\documentclass[12pt]{iopart}

\usepackage{graphicx}
\begin{document}

\title{A case study for terahertz-assisted single attosecond pulse generation}

\author{Emeric Balogh$^1$, Katalin Kov\'acs$^{1,2}$, Valer To\c sa$^2$ and Katalin Varj\'u$^1$}
\address{$^1$Department of Optics and Quantum Electronics, University of Szeged, D\'om t\'er 9, Szeged, HU-6701, Hungary}
\address{$^2$National Institute for R\&D of Isotopic and Molecular Technologies, Str. Donath 65-103, Cluj-Napoca, RO-400293, Romania}
\ead{ebalogh@titan.physx.u-szeged.hu}

\begin{abstract}
We numerically investigate the use of strong THz radiation in assisting single attosecond pulse generation by few-cycle, 800 nm laser pulses. We optimize focusing conditions to generate short and powerful single attosecond pulses of high energy photons by keeping the parameters of the THz field within the limits achieved experimentally. We show that using optimal focusing geometry isolated attosecond pulses shorter than 100 as can be obtained even in the absence of further gating or XUV compression techniques, using an 8 fs generating pulse. Furthermore, quantum path control of short and long trajectory components is demonstrated by varying the delay between the THz and IR pulses.
\end{abstract}


\section{Introduction and motivation}
Recent developments in THz field generation resulted in the production of extreme high electric fields exceeding 100 MV/cm with a carrier frequency up to 72 THz and stable carrier-to-envelope phase (CEP) \cite{sell}.
This field strength is already comparable with the peak electric field of laser pulses used for high-order harmonic generation (HHG) in noble gases (usually between 300 and 1000 MV/cm).
Since THz fields with the highest intensities are produced by difference frequency generation from amplified laser pulses,
the resulting THz pulses are naturally synchronized in time with the laser pulses \cite{sell}.

Single atom calculations
have already revealed several aspects of HHG assisted by THz or static electric fields.
The production of even harmonics and the increased emission rate in the lower plateau region has been demonstrated when multi-cycle infrared (IR) laser pulses are used as driving field \cite{minqi}.
With higher intensity THz or static electric fields the extension of the cutoff has been observed producing a double-plateau-structured spectrum \cite{wang1998,hong,Taranukhin}.
Using few-cycle laser pulses, the addition of the THz field can create a super-continuum in the spectra enabling the production of single attosecond pulses (SAP) \cite{songsong}.
With chirped IR driving pulses the created supercontinuum, theoretically, can support 10 attosecond short SAPs \cite{yang}.
In calculations predicting the production of SAPs however the used amplitude of the THz or static electric field has been higher than what is achievable experimentally, and the used laser pulses were at most 6 fs long \cite{songsong,yang}.
The experimentally obtained 100 MV/cm electric field requires very tight focusing of the THz beam.

Here we present the effects of THz field on HHG in gases beyond the single atom level, by modeling the process in a realistic focusing geometry, and using parameters of the THz pulse obtained experimentally by Sell et al. \cite{sell}.
In a recent study \cite{mPRA2011} it was shown that in the presence of a THz field macroscopic effects can help the selection of a SAP from the attosecond pulse train when using 8 fs driving pulses.
In the present study we discuss the importance of focusing geometry on phase matching and harmonic yield, and show that despite the limited THz pulse energy the most powerful SAP can be produced by relatively loose focusing and lower THz electric field strength.
We also show that the selection of the short or long trajectory components can be achieved by varying the delay between the THz and IR pulses.

Our goal is to model experimental conditions as close as possible, therefore a full three dimensional model has been used which allows the optimization of parameters that can be measured and controlled experimentally\cite{tosamodel}.
For calculating the single atom response the established theory of strong-field approximation is used, known to well reproduce the important aspects of HHG in gases\cite{lewenstein}.
The propagation equations for the laser, THz, and harmonic fields are solved by using a Fourier transform method in paraxial approximation \cite{priori}, by taking into account the effects of absorption, dispersion on atoms and on electrons, and the optical Kerr effect.
The three fields propagate independently in a medium with refractive indices mainly influenced by the plasma created by the total electric field.

\section{Results}
\subsection{Basic configuration}
We have modeled harmonic generation in neon gas by 8 fs, 800 nm infrared (IR) laser pulse combined with a THz pulse having 19 $\mu$J energy and 72 THz carrier frequency (4.17 $\mu$m wavelength) focused at 31 $\mu$m spot size producing 108 MV/cm peak field strength as generated and measured by Sell et al. \cite{sell}.
The pulses have been assumed to be Gaussian both in space and time, the THz pulse having 76 fs duration to be consistent with the above mentioned parameters.
The beam diameters are assumed to be 5 mm and 26 mm for the IR and THz beams, so that using an optical element with f=0.6 m focal distance yields the same w$_0$=31 $\mu$m beam waist for both.
The two fields have collinear polarization and propagate in the same direction, while the gas medium containing neon at 33 mbar pressure is placed right after the focus.
The wavelength, energy and duration of the THz pulse were consistent with the experimentally demonstrated values through all the calculations, while the energy of the IR pulse (usually not a limiting factor in experiments) has been chosen to produce 6$\times$10$^{14}$ W/cm$^2$ intensity in the focus for all cases.

In order to optimize the generating conditions the effects of several parameters were studied.
The focusing geometry was optimized, limited by the constraints of using maximum spatial overlap between the IR and THz pulses (i.e. same beam waist).
Besides the focusing conditions the delay between the two pulses, the length and position of the gas cell were optimized in order to generate short and powerful SAPs of high energy photons.

\begin{figure}[htb]
\centerline{\includegraphics{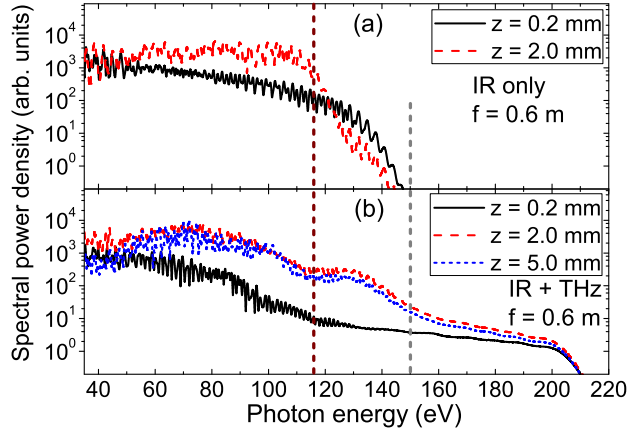}}
\caption{Spectral power density of the propagated harmonic field at the exit of a 0.2 mm (black solid lines) and 2.0 mm (red, dashed lines) gas cell for the cases when only the IR field (a) and when the combined fields (b) are focused by a f = 0.6 m mirror, producing a beam waist of 31 $\mu$m. The blue dotted line on graph (b) represents the same quantity for a 5 mm cell showing the decreased yield due to phase mismatch in the tight focusing conditions. The vertical lines at 116 and 150 eV show the lower limit of spectral filtering used to synthesize the attosecond pulses.}
\label{fig1}
\end{figure}

Our calculations show that indeed the cutoff of the harmonic radiation can be extended from 110 eV to 200 eV in the near field (see Fig.\ref{fig1}) when an 8 fs, 78 $\mu$J IR pulse is used in the above mentioned focusing geometry in a 2 mm long target cell.
However, the tight focusing used to obtain the extreme high electric field of the THz pulse results a Rayleigh range of just 0.7 mm (3.7 mm for the IR beam), which is not beneficial to phase match the generated harmonics as illustrated in Fig.\ref{fig1}.b showing decreasing signal after 2 mm propagation.
By increasing the beam waist the peak intensity of the THz field drops, limiting the achievable cutoff, but helping to phase match the higher spectral components.

\begin{figure}[htb]
\centerline{\includegraphics{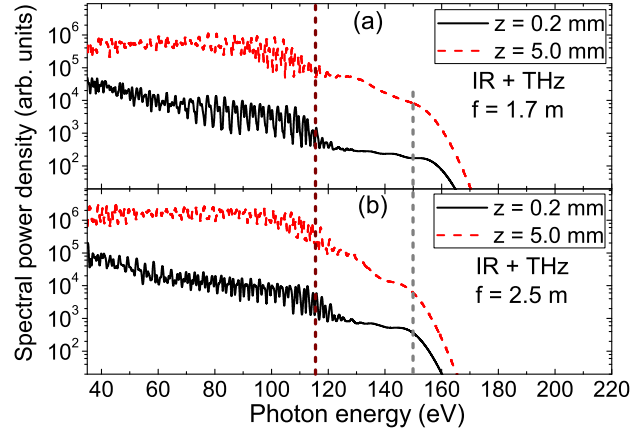}}
\caption{Spectral power density of the propagated harmonic field at the exit of a 0.2 mm (black solid lines) and 5.0 mm (red, dashed lines) gas cell when both the IR field and THz fields are focused by a f = 1.7 m (a), and by a f = 2.5 m mirror (b) producing beam waists of 85 and 125 $\mu$m respectively. The vertical lines at 116 and 150 eV show the lower limit of spectral filtering used to synthesize the attosecond pulses.}
\label{fig2}
\end{figure}

\subsection{Single attosecond pulses}

\begin{figure}[htb]
\centerline{\includegraphics{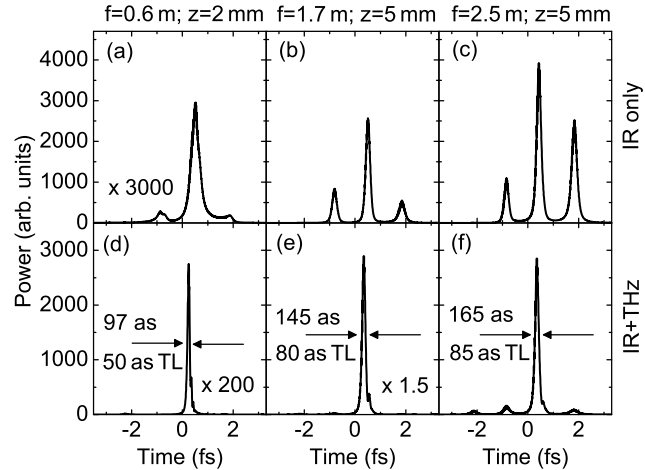}}
\caption{Attosecond pulses obtained by selecting harmonic radiation above 116 eV (harmonic 75) using different focusing geometries with the IR only (top row), and with the combined field (bottom row). The transform limit of the pulses is also shown in cases when SAP is obtained.}
\label{fig3}
\end{figure}

Using f=1.7 m focusing and 0.585 mJ IR pulse energy the corresponding beam waist is 85 $\mu$m; the addition of the THz field with 38 MV/cm peak amplitude and same 85 $\mu$m waist extends the cutoff by 30 eV (see Fig.\ref{fig2}.a), reaching 154 eV compared to the cutoff at 124 eV obtained using only the IR pulse.
The plateau region is phase matched during propagation through a 5 mm gas cell.
This configuration supports the production of 145 as SAP obtained by selecting harmonics above 116 eV (see Fig.\ref{fig3}.e).
The larger interaction volume resulting from the increased spot size and favorable phase matching conditions through a 5 mm cell result in an increase of more than two orders of magnitude in the peak power of the generated SAP (see Fig.\ref{fig3} d,e,f), despite the lower cutoff (i.e. narrower bandwidth).

Further loosening the focusing geometry (f = 2.5 m, 125 $\mu$m beam waist, 1.25 mJ IR pulse energy) the amplitude of the THz field drops to 27 MV/cm and the cutoff is extended by only 20 eV (Fig2.b) compared to the IR only case with a cutoff at 124 eV.
Using the same spectral filtering ($\geq$116 eV), satellite pulses appear around the main pulse, and the peak power of the main pulse does not increase significantly (see Fig.\ref{fig3}.f and  Table.\ref{table1}) despite the larger interaction volume.

\begin{figure}[htb]
\centerline{\includegraphics{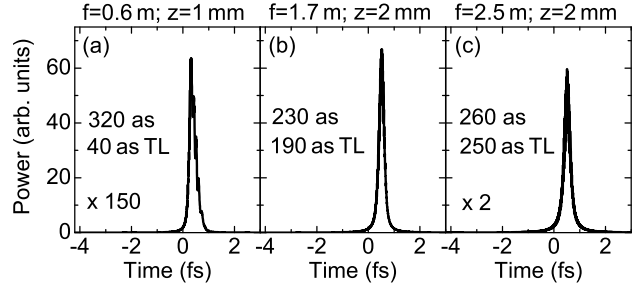}}
\caption{Attosecond pulses obtained by selecting harmonic radiation above 150 eV (harmonic 97) using different focusing geometries with the combined field. The transform limit of the pulses is also shown.}
\label{fig4}
\end{figure}

Our analysis shows that the production of 97 as SAP is mainly attributed to the elimination of long trajectory components by phase mismatch.
However, the elimination of the two satellite pulses observable in the IR only case (half optical cycle before and after the main pulse, see top row of Fig.\ref{fig3}) can be attributed to the presence of the THz electric field breaking the inversion symmetry of the system.
This lowers the cutoff at every second half cycles of the IR driving field \cite{hong, wang1998}, falling below the lower limit of the used spectral filtering.
These pulses appear on Fig.\ref{fig3}.f due to the decreased THz field strength.

By increasing the lower limit of the spectral filtering SAP still can be obtained at the cost of reduced power.
Selecting only harmonics above 150 eV (harmonic 97) SAP can be obtained in all the focusing geometries used so far (see Fig.\ref{fig4} and Table.\ref{table1}).
However, the loose focusing (f=2.5 m) also shifts the cutoff below 150 eV, resulting a reduced pulse power compared to the case with f=1.7 m focusing.
We note that the SAP obtained this way is nearly transform limited (Fig.\ref{fig4}.c), because cutoff harmonics possess no chirp \cite{krausz} and the bandwidth is very narrow.

The 50 as transform limit of the SAP presented in Fig.\ref{fig3}.d corresponds to an effective bandwidth of 36 eV (with a time-bandwidth product of 0.44, characteristic of Gaussian pulses), which can be explained by the strong drop of the harmonic yield at $\approx$145 eV seen in Fig.\ref{fig1}.b.
By selecting only harmonics above 150 eV the spectrum is flatter which explains the shorter transform limit of the synthesized SAP shown in Fig.\ref{fig4}.a.

The differences in the degree of cutoff extension by the THz field at different focusing geometries can be attributed not only to the different amplitude of the THz field, but to macroscopic effects as well \cite{mPRA2011}.
Using saddle point analysis we have found that at single atom level a long wavelength THz (compared to the 800 nm IR) or static electric field increases the harmonic radiation's cutoff by 0.4-1 eV for each MV/cm field strength.
Although this number also depends on the intensity and wavelength of the IR (it is $\approx$ 0.65 eV/(MV/cm) for 800 nm IR having 6$\times$10$^{14}$ W/cm$^2$ peak intensity), and the scaling with the THz or DC field amplitude is not exactly linear.
For a more detailed analysis of the mechanism of cutoff extension see for example the paper by Taranukhin et al. \cite{Taranukhin}.

\begin{table*}[htb]
  \centering
  \caption{Summary of the THz field and SAP parameters obtained for different focusing geometries using two different spectral filters, and cell lengths optimized for SAP peak power.
  Contrast ratio labeled $\infty$ means that the contrast is higher than the precision of our calculations ($\approx$10$^{7}$).}
  \begin{tabular}{cccccccc} \\ \hline
    f    & w$_0$    & E$_{peak}$ & L$_{optimal}$ & cutoff & SAP duration       & SAP peak power       & contrast         \\
    -    & -        & -          & 116/150 eV*   & -      & 116/150 eV         & 116/150 eV           & 116/150 eV       \\
    (m)  & ($\mu$m) & (MV/cm)    & (mm)          & (eV)   & (as)               & (arb. units)         & (dB)             \\ \hline
    0.61 & 31       & 108        & 2.1 / 0.6     & 205    & \textbf{97} / 136  & 14.2 / 0.62          & 22.3 / 34.4      \\
    1.1  & 56       & 60         & 5.1 / 1.6     & 163    & 110 / 184          & 384 / 24.1           & 22.1 / $\infty$  \\
    1.7  & 86       & 38         & 5.2 / 2.3     & 155    & 150 / 230          & 1947 / \textbf{68.2} & 21.6 / $\infty$  \\
    2.45 & 125      & 27         & 4.3 / 2.1     & 148    & 174 / 260          & \textbf{3085} / 29.8 & 12.2 / $\infty$  \\ \hline
  \end{tabular}
  \begin{tabular}{l}
   *116/150 eV stands for distinguishing the two spectral filters
  \end{tabular}
  \label{table1}
\end{table*}

A summary of the results is presented in Table.\ref{table1}.
The values for SAP duration, peak power, and contrast presented in the table are calculated for the optimal cell lengths in the actual focusing geometry, whereas the values presented in figures \ref{fig3} and \ref{fig4} are calculated for cell lengths more commonly used in experiments.
The contrast ratio is defined between the peak powers of the two most powerful attosecond pulses.

For the position of the gas cell we have found that the most reliable solution is to place it with the input pinhole in the focus.
With the looser focusing geometries slightly better results were obtained by moving the cell 1 mm before the focus, however the increase is not significant.
With the f=1.7 m case for example 10\% increase can be obtained in the SAP peak power when an extended, 6 mm long cell is placed beginning 1 mm before the focus.

\begin{figure}[htb]
\centerline{\includegraphics{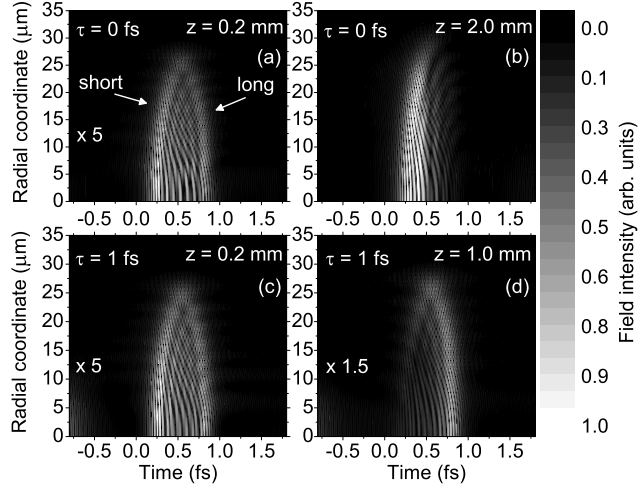}}
\caption{Intensity maps in (r,t) of the propagated harmonic field at different axial (z) coordinates. Top row shows the selection of short trajectories by phase matching after 2 mm propagation in case of 0 delay, while the bottom row shows the selection of long trajectories by phase matching after 1 mm propagation obtained by delaying the THz field by 1 fs compared to the IR. In this case f=1.1 m focusing is used producing 56 $\mu$m beam waist.}
\label{fig5}
\end{figure}

\subsection{Quantum path control}

So far the two pulses were assumed to be synchronized, i.e. to have a field maximum at t=0.
In this case the short trajectory components are phase matched and the corresponding long ones are eliminated due to phase mismatch.
To demonstrate the effect, the intensity of the propagated XUV pulse is plotted in (r,t) maps (see Fig.\ref{fig5}).
First we calculate the radiation produced in a very short cell (0.2 mm, see Fig.\ref{fig5}.a), where macroscopic effects do not start to play, and such illustrating the single atom results. On axis (r=0) we observe almost half a cycle of modulated radiation consisting of both short and long trajectory components \cite{lewenstein,mPRA2011} generating earlier (short) and later (long) emissions.
The short- and long-trajectory components merge into the cutoff with increasing radial coordinate corresponding to decreasing IR and THz field strength.
Our analysis (not illustrated here) shows that both short and long trajectory components are generated at any z coordinate in the cell, however the long ones gradually disappear from the propagated field due to phase mismatch (see Fig.\ref{fig5}.b). We find that the short trajectories dominate the harmonic radiation after 1 mm propagation (not shown), with the long ones completely eliminated after 2 mm.
Using the same f=1.1 m focusing mirror, (56 $\mu$m beam waist) and delaying the THz pulse by 1 fs compared to the IR, we observe an almost identical field intensity map at the beginning of the cell (see Fig.\ref{fig5}.c). However, during propagation only
long trajectories are phase matched in the first mm of the gas cell, with the corresponding short ones eliminated (see Fig.\ref{fig5} bottom row).
We conclude that by using well chosen focusing and varying the delay between the THz and IR pulses, the selection of long trajectories can be achieved allowing one to select the sign of the chirp of the resulting SAP.

Another effect reported in single atom calculations is the increase of the emission rate in the lower part of the plateau.
We have also observed this effect, however in the best of our cases the harmonic yield increased by only several times compared to the IR only case, not by orders of magnitude as reported in \cite{wang1998}.

\section{Conclusion}
In conclusion we have analyzed the effects of experimentally obtained THz pulses on HHG by modeling the process using a complete three-dimensional model.
The importance of focusing geometry on phase matching and harmonic yield has been discussed and shown that, despite the limited THz pulse energy, the most powerful SAP can be produced by relatively loose focusing.
Short, isolated attosecond pulse having a duration of 97 as is predicted using tight focusing and 2 mm long gas cell.
We have also shown that the selection of the short or long trajectory components (defining the sign of the resulting SAP's chirp) can be achieved by varying the delay between the THz and IR pulses.

\section*{Acknowledgement}
The project was supported by the European Community's FP7 Programme under contract ITN-2008-238362 (ATTOFEL).
The project was partially funded by "TAMOP-4.2.1/B-09/1/KONV-2010-0005 - Creating the Center of Excellence at the University of Szeged" supported by the European Union and co-financed by the European Social Fund. KK acknowledges support from CNCS-UEFISCDI project no. PN-II-RU-PD-2011-3-0236.
VT acknowledges support from the ESF activity SILMI.
KV acknowledges support from the Bolyai Foundation.
We are grateful to NIIF Institute and NIRDIMT Data Center for computing time in their supercomputing centers.


\section*{References}
\begin{thebibliography}{99}
\bibitem{sell} Sell A, Leitenstorfer A and Huber R 2008
{\it Opt. Lett.} \textbf{33} 2767--9

\bibitem{minqi} Bao M Q and Starace A F 1996
{\it \PR}A \textbf{53} R3723--26

\bibitem{wang1998}
Wang B, Li X and Fu P 1998 {\it \jpb}
 \textbf{31} 1961

\bibitem{hong}
Hong W, Lu P, Lan P, Zhang Q and Wang X 2009
{\it Opt. Express} \textbf{17} 5139--46

\bibitem{Taranukhin}
Taranukhin V D and Shubin N Y 2000
{\it \JOSA} B \textbf{17} 1509--16

\bibitem{songsong}
Tang S, Zheng L and Chen X 2010
{\it Optics Communications}
  \textbf{283} 155--9

\bibitem{yang}
Xiang Y, Niu Y and Gong S 2009
  {\it \PR}A \textbf{79} 053419

\bibitem{mPRA2011}
Balogh E, Kov\'acs K, Dombi P, F\"ul\"op J A, Farkas G, Hebling J, To\c sa V  and Varj\'u K 2011
{\it \PR}A \textbf{84} 023806

\bibitem{tosamodel}
Takahashi E, Tosa V, Nabekawa Y and Midorikawa K 2003 
{\it \PR}A \textbf{68}  023808

\bibitem{lewenstein}
Lewenstein M, Balcou P, Ivanov M Y, L'Huillier A and Corkum P B 1994
{\it \PR}A \textbf{49} 2117--32

\bibitem{priori}
Priori E, Cerullo G, Nisoli M, Stagira S, De~Silvestri S, Villoresi P,  Poletto L, Ceccherini P, Altucci C, Bruzzese R and de~Lisio C 2000
{\it \PR}A \textbf{61} 063801

\bibitem{krausz}
Krausz F and Ivanov M 2009
 {\it \RMP} \textbf{81} 163--234
    
\end{thebibliography}
\end{document}